\documentclass[journal=acscii,manuscript=article]{achemso}

\usepackage[version=3]{mhchem} 
\usepackage[T1]{fontenc}       

\usepackage{amssymb, amsmath,amsfonts}
\usepackage{graphicx}
\usepackage{dcolumn}
\usepackage{bm}
\usepackage{multirow}
\usepackage{xcolor}
\usepackage[percent]{overpic}

\title{Tension-dependent Free Energies of Nucleosome Unwrapping}

\author{Joshua Lequieu}
\affiliation{Institute for Molecular Engineering\\ University of Chicago, Chicago, IL 60637 USA}
\author{Andr\'es C\'ordoba}
\affiliation{Institute for Molecular Engineering\\ University of Chicago, Chicago, IL 60637 USA}
\author{David C. Schwartz}
\affiliation{Laboratory for Molecular and Computational Genomics, Department of Chemistry, Laboratory of Genetics, and UW-Biotechnology Center, University of Wisconsin-Madison, Madison, WI 53706, USA}
\author{Juan J. de Pablo}
\affiliation{Institute for Molecular Engineering\\ University of Chicago, Chicago, IL 60637 USA}
\alsoaffiliation{Materials Science Division\\ Argonne National Laboratory, Argonne, IL 60439 USA}
\email{depablo@uchicago.edu}

\date{\today}

\begin{document}

\begin{abstract}
Nucleosomes form the basic unit of compaction within eukaryotic genomes and their locations represent an important, yet poorly understood, mechanism of genetic regulation. Quantifying the strength of interactions within the nucleosome is a central problem in biophysics and is critical to understanding how nucleosome positions influence gene expression. By comparing to single-molecule experiments, we demonstrate that a coarse-grained molecular model of the nucleosome can reproduce key aspects of nucleosome unwrapping. Using detailed simulations of DNA and histone proteins, we calculate the tension-dependent free energy surface corresponding to the unwrapping process. The model reproduces \textit{quantitatively} the forces required to unwrap the nucleosome, and reveals the role played by electrostatic interactions during this process. We then demonstrate that histone modifications and DNA sequence can have significant effects on the energies of nucleosome formation. Most notably, we show that histone tails are crucial for stabilizing the outer turn of nucleosomal DNA.
\end{abstract}
\maketitle

\section*{Introduction}

Eukaryotic genomes are packaged into a compact, yet dynamic, structure known as chromatin.
The basic building block of chromatin is the nucleosome, a disk-like structure consisting of 147 base pairs of DNA wrapped into 1.7 superhelical turns around proteins known as histones\cite{Luger1997,Davey2002}.
These histone proteins form what is known as the histone octamer, a stable protein complex consisting of two copies of histone proteins H2A, H2B, H3 and H4.
The surface of the histone octamer is highly positive, which interacts favorably with the negative backbone of DNA.
As a result, at sufficiently low ionic conditions, nucleosomes are stable and spontaneously form.

The locations of nucleosomes along the genome play a central role in eukaryotic regulation. 
DNA segments incorporated into nucleosomes are inaccessible to other DNA binding proteins, including transcription factors and polymerases, and thus nucleosomes must be disrupted in order for the cellular machinery to access nucleosomal DNA.
As such, the positions occupied by nucleosomes provide an additional, important mechanism by which eukaryotic genomes are regulated.
Indeed, past work has demonstrated that deregulation of these processes is implicated in numerous diseases, including cancer\cite{Hendrich2001,Bhaumik2007,Sadri-Vakili2007}.
Quantifying the strength of interactions within the nucleosome structure and the forces required to disrupt them is of fundamental importance to understanding the delicate dynamics of chromatin compaction.

Optical-trapping single-molecule techniques have been particularly effective at probing multiple interactions that underlie the nucleosome.
In these experiments, chromatin fibers \cite{Brower-Toland2002, cui2000pulling,Gemmen2005,Brower-Toland2005} or single nucleosomes \cite{Mihardja2006, Kruithof2009,Hall2009,Sheinin2013,Mack2012,Ngo2015} are subjected to pico-newton scale forces, thereby providing the ability to precisely perturb the native nucleosome structure.
By analyzing the deformations that result from these forces, one can infer the underlying strength of binding energies within the nucleosome.
Following the initial work by Mihardja \textit{et al.}~\cite{Mihardja2006}, a consensus is emerging \cite{Kruithof2009,Sheinin2013,Ngo2015} in which a single nucleosome is disrupted in two stages. In the first, at 3 pN, the outer wrap of DNA is removed from the histone surface.
This first wrap is removed gradually and is considered to be an equilibrium process, where spontaneous unwrapping and rewrapping events can be observed under a constant force.
The second transition occurs at forces 8-9 pN and occurs rapidly via so-called ``rips'', where the remaining wrap of DNA is suddenly released.
More recently, these transitions have been shown to depend on torque (i.e. DNA supercoiling via twist) \cite{Sheinin2013}, and to occur antisymmetrically due to variability in the bound DNA sequence \cite{Ngo2015}.

Several theoretical and computational studies have sought to help interpret these experimental results.
Following the initial work of Kuli{\'{c}} and Schiessel\cite{Kulic2004}, most current treatments represent the nucleosome as an oriented spool, and the unbound DNA as a semiflexible worm-like chain \cite{Sudhanshu2011,Mollazadeh-Beidokhti2012}.
While earlier studies were only able to detect a single distinct unwrapping transition \cite{Kulic2004,wocjan2009brownian}, consistent with the first experiments \cite{Brower-Toland2002}, more recent work\cite{Sudhanshu2011,Mollazadeh-Beidokhti2012} has been able to reproduce the two transitions observed by Mihardja \textit{et al.}~\cite{Mihardja2006}.
By relying on simple, primarily analytic models, these studies have provided significant insights into the fundamental physics that govern interactions within the nucleosome. Such approaches, however, have necessarily had to invoke assumptions and introduce adjustable parameters in order to describe experiments
\cite{Sudhanshu2011,Mollazadeh-Beidokhti2012} (e.g. the DNA-histone binding energy).
This limits their ability to predict nucleosomal behavior under different conditions, such as variations in DNA sequence or ionic environment, without resorting to additional experimental data.
Additionally, these models cannot explicitly account for histone modifications, which are central to nucleosome positioning and higher-order chromatin structure\cite{Jenuwein2001,kurdistani2004mapping,barski2007high,zhou2011charting,Brower-Toland2005}.

A complementary approach, which should in principle enable prediction of nucleosomal interactions under a wide array of situations, could rely on molecular models where the nucleosome can be assembled or disassembled explicitly.
Though these approaches are particularly promising, their success has been frustrated by the inability to access the experimentally relevant time scales of stretching, typically rates of ~100nm/second.
Clearly, these time scales are inaccessible to atomistic simulations, yet even a highly coarse-grained spool-like model of the nucleosome only achieved stretching rates several orders of magnitude too fast\cite{wocjan2009brownian}.
There is therefore a need to develop models and methodologies to facilitate more direct comparisons between optical-trapping experiments and molecular-level calculations. If successful, such models could reveal the subtle, yet incredibly important effects of DNA-sequence and histone modifications on nucleosome stability.


In this work, we build on a recently proposed coarse-grained model of the nucleosome\cite{freeman2014coarse,Freeman2014a,Li2011} to examine its response to external perturbations.
A computational framework is proposed in which the tension-dependent response of the nucleosome is examined at equilibrium, thereby providing access to the free energy of nucleosome unwrapping under tension.
Our results are found to be in agreement with experimental measurements by Mihardja \textit{et al.}~\cite{Mihardja2006}, and serve to demonstrate that it is indeed possible to reproduce the absolute binding free energies of nucleosome formation in terms of purely molecular-level information, without resorting to adjustable parameters.
Importantly, that model is then used to predict the impact of DNA sequence and histone modifications on the \textit{relative} free energies of binding.

\section*{Results}\label{sec:results}

\begin{figure}
    \includegraphics[width=3.33in]{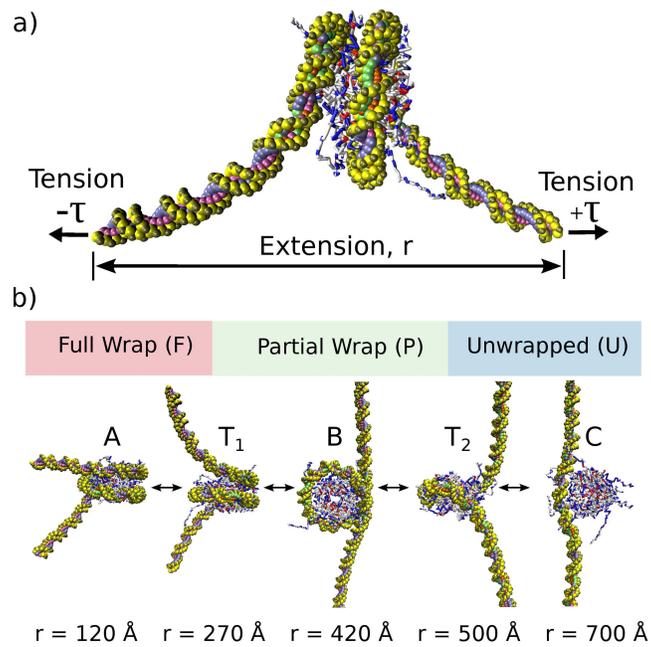}
    \caption{%
    Model of Nucleosome Unwrapping.
    a) Coarse-grained topology of Nucleosome. DNA is represented by 3SPN2.C\cite{freeman2014coarse}, the histone proteins by AICG\cite{Li2011}. Both the end-to-end extension, $r$, and tension, $\tau$, are constrained during a simulation.
    b) Unwrapping process. During extension, the wraps of DNA around histone proteins are removed one by one. $T_1$ and $T_2$ denote the transition states separating the first ($A \leftrightarrow B$) and second ($B \leftrightarrow C$) unwrapping events.
    Figures were generated using VMD\cite{Humphrey1996}.
    }
    \label{fig:pull-snapshots}
\end{figure}

A schematic of our simulation setup is shown in Figure~\ref{fig:pull-snapshots}a.
As with optical-trapping experiments, the ``state'' of the nucleosome is represented by two parameters: the tension (or force) exerted on the DNA molecule, $\tau$, and the extension of the DNA ends, $r$.
To facilitate comparison with experiments \cite{Mihardja2006}, the ends of the DNA are not torsionally constrained.
Figure~\ref{fig:pull-snapshots}b shows instantaneous configurations of the nucleosome model for five different values of extension, $r$.
Consistent with previous observations, the outer wrap of the nucleosome is first removed ($A \rightarrow T_1 \rightarrow B$), followed by the inner wrap ($B \rightarrow T_2 \rightarrow C$).

\begin{figure}
    \includegraphics[width=3.33in]{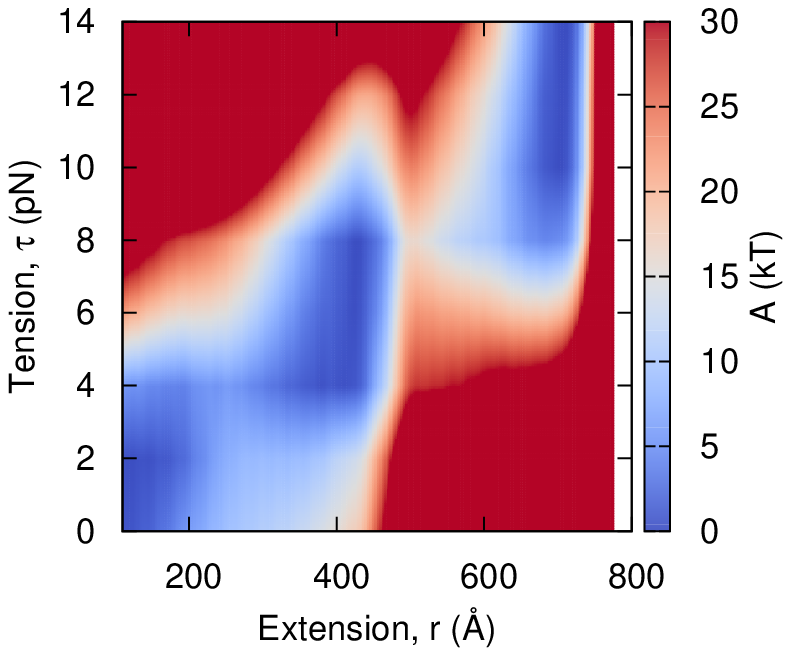}
    \caption{%
    Tension-dependent free energy surface of nucleosome unwrapping for 601 positioning sequence.
    The free energy surface demonstrates minima at extensions of $r \approx 120 \AA, \approx 420, \approx 700 \AA$, depending on tension.
    As tension increases, the minimum-energy extension shifts to larger values.
    Consistent with Mihardja \textit{et al.}~\cite{Mihardja2006}, two transitions are observed.
    }
    \label{fig:tension-sep}
\end{figure}

In order to quantify these transitions, we examine the tension-dependent free energy of nucleosome unwrapping.
By calculating the tension-dependent free energy instead of a simple force-extension curve, as done previously \cite{wocjan2009brownian,Kenzaki2015}, we can determine the true equilibrium behavior of the unwrapping process.
Additionally, by performing simulations at equilibrium, we circumvent the issue of time scales that frustrate comparisons of traditional non-equilibrium molecular simulations to optical pulling experiments.

A representative two-dimensional tension-extension free energy surface for the 601 positioning sequence\cite{Lowary1998} is shown in Figure~\ref{fig:tension-sep}.
Rather than increasing linearly with tension, the extension is quantized into three well defined vertical bands, located at $\approx 120\AA$, $420\AA$, and $700\AA$, corresponding to states ``A'', ``B'' and ``C'' in Figure \ref{fig:pull-snapshots}.
At low tension ($\tau < 3pN$), a low extension ($r<200\AA$) is preferred.
As tension is increased ($\tau \approx 4-8 pN$), the minimum free energy shifts to intermediate values of extension ($r \sim 420\AA$).
At higher tension ($\tau > 8pN$), the minimum free energy shifts to larger values of extension ($r = 700\AA$).
The free energy penalty of low tension and high extension (e.g. $\tau = 3, r = 700$) or high tension with low extension (e.g. $\tau = 12, r =200$) results in large energy barriers $>40\,kT$.

\begin{figure}
    \includegraphics[width=3.33in]{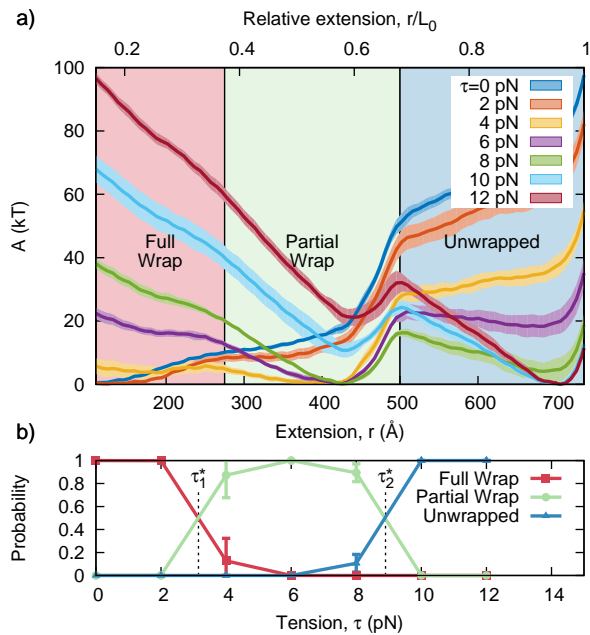}
    \caption{%
    a) Free energy versus extension for different values of tension with the 601 positioning sequence.
    Based on the locations of the transition states, $T_1$ and $T_2$, three basins can be defined: ``Fully Wrapped'', ``Partially Wrapped'', and ``Unwrapped''.
    $L_0$ represents the contour length of the DNA molecule.
    b) Probability of observing the nucleosome in each free energy basin for different tensions.
    The ``Fully'' and ``Partially'' Wrapped states are at equilibrium (i.e. equal probability) when $\tau_1^* = 3.2$ pN. The ``Partially'' and ``Unwrapped'' states are at equilibrium when $\tau_2^* = 8.9$ pN.
    Error bars represent standard deviation across four independent simulations.
    }
    \label{fig:fes-force}
\end{figure}

The tension-dependent transition can also be visualized by plotting one-dimensional ``slices'' of the free energy surface at different values of tension (Figure~\ref{fig:fes-force}a).
Visualizing the data in this way clearly demonstrates that there are three basins of nucleosome extension: ``Fully Wrapped'', ``Partially Wrapped'' and ``Unwrapped''.
The basin that is favored depends on the tension applied to the DNA ends.
As tension increases, the free energy minima shifts first from the ``Fully Wrapped'' to the ``Partially Wrapped''  basin, and then to the ``Unwrapped'' basin.
The boundaries of these basins are defined by the locations of the transition states (i.e. local maximum in the free energy) that separate neighboring basins.
The transition states separating the $A \rightarrow B$ transition, $T_1$, and the $B \rightarrow C$ transition, $T_2$, are shown in Figure~\ref{fig:pull-snapshots}b.

Once these three basins are defined, we can determine the precise tension at which the outer and inner DNA turns unwrap from the nucleosome.
This is obtained by converting the tension-dependent free energy into probabilities, and then integrating these probabilities to determine the total probability of finding the system in each basin (see Materials and Methods). 
The corresponding results are shown in Figure~\ref{fig:fes-force}b; it can be appreciated
that the probability of finding the system in the ``Fully Wrapped'' or ``Partially Wrapped'' basin is equivalent when $\tau \approx 3.2$ pN.
Thus, when $\tau \approx 3.2$ pN the outer turn of nucleosomal DNA is in equilibrium (in a statistical mechanics sense) with its unbound state.
We define this tension as $\tau_1^*$.
Similarly, the probability of the nucleosome in the ``Partially Wrapped'' and ``Unwrapped'' basins is the same (i.e. the inner wrap is in equilibrium when $\tau \approx 8.9$ pN, defined as $\tau_2^*$).
These values are in quantitative agreement with those measured by Mihardja \textit{et al.}~\cite{Mihardja2006}, who observed that the outer and inner DNA loops were removed at 3 pN and 8-9 pN, respectively.

\begin{figure}
    \includegraphics[width=3.33in]{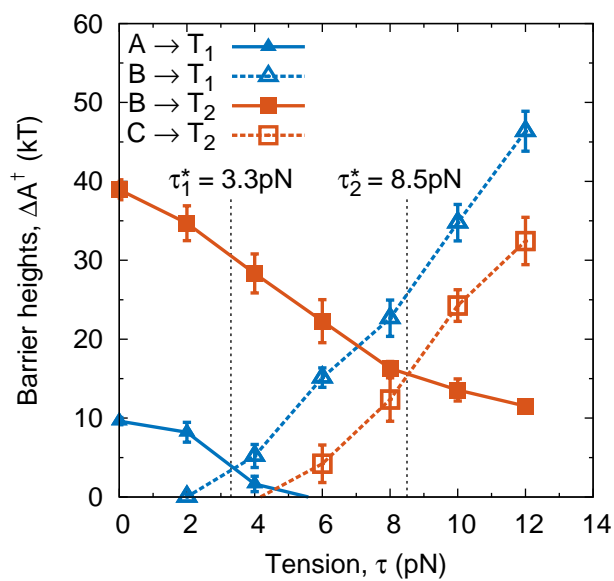}
    \caption{%
    Free energy barrier heights of nucleosome unwrapping for 601 positioning sequence. Solid lines represent the unwrapping (forward) reactions, dotted lines represent wrapping (reverse) reactions.
    When the unwrapping and wrapping barriers are equal, the two basins are at equilibrium with one another.
    This is found when $\tau_1^* = 3.3$ pN for the outer wrap, and $\tau_2^* = 8.5$ pN for the inner wrap.
    $\Delta A^\dagger (\tau_1^*) = 4\,kT$ and $\Delta A^\dagger (\tau_2^*) = 16\,kT$.
    Error bars represent standard deviation across four independent simulations.
    }
    \label{fig:barrier}
\end{figure}

A complementary approach to estimate $\tau_1^*$ and $\tau_2^*$, is to determine the tension at which the free energy barriers of the forward and reverse reactions are equal \cite{Sudhanshu2011}.
Figure \ref{fig:barrier} shows the corresponding tension-dependent free energy barriers of the outer ($A \leftrightarrow T_1 \leftrightarrow B$) and inner unwrapping ($B \leftrightarrow T_2 \leftrightarrow C$) events.
At low tension, the energy barriers for the forward reactions, $A \rightarrow B$ and $B \rightarrow C$, dominate and the forward (i.e. unwrapping) reaction rate is low.
As tension increases, the energy barriers for the forward reactions decrease, while those for the reverse increase, thereby causing the unwrapping reaction to proceed at a higher rate.
When the energy barriers of the forward and reverse reactions are equal, the two basins are at equilibrium (in a transition state theory sense) and $\tau_1^*$ and $\tau_2^*$ can be determined.
These unwrapping tensions are estimated to be 3.3 pN and 8.5 pN, in excellent agreement with the probability-based analysis of Figure~\ref{fig:fes-force}b.

The magnitude of the free energy barrier also helps explain the observation by Mihardja \textit{et al.}~\cite{Mihardja2006} that the outer turn of DNA can be removed reversibly, while the inner turn cannot.
Since the energy barrier separating the ``Fully'' and ``Partially'' wrapped states is only $\approx5\,kT$, the system can quickly transition between states when held at $\tau = \tau_1^*$.
In contrast, the ``Partial Wrap'' and ``Unwrapped'' states are separated by an energy barrier of $\approx18\,kT$, indicating that even at equilibrium the $P \leftrightarrow U$ transition occurs slowly.
Thus, removal of the outer wrap may appear to be reversible on the time scales of a typical optical trapping experiment, while the inner wrap may not.
Further, because force-extension curves are usually obtained via optical trapping by pulling a nucleosome at a fixed velocity, the experiments may not observe a $P \rightarrow U$ transition until $\tau > \tau_2^*$.
This would cause the experiments to overestimate the value of $\tau_2^*$, and lead to a sudden, irreversible ``ripping'' event.
We also note that the barrier estimates in this work ($\Delta A^\dagger_1 = 4\,kT, \Delta A^\dagger_2 = 16\,kT$) are in excellent agreement to those predicted by Sudhanshu \textit{et al.}~\cite{Sudhanshu2011} ($\Delta A^\dagger_1 \approx 6\,kT, \Delta A^\dagger_2 \approx 15\,kT$).

\subsection*{Electrostatics, Sequence Dependence, Histone Modifications}

Having validated the proposed model against experimental data \cite{Mihardja2006}, we now examine the influence of ionic environment, DNA sequence, and histone modifications on the stability of the nucleosome. Such variations can have a significant impact on nucleosome formation, and the precise molecular origins of their impact is still poorly understood.

\begin{figure}
    \includegraphics[width=3.33in]{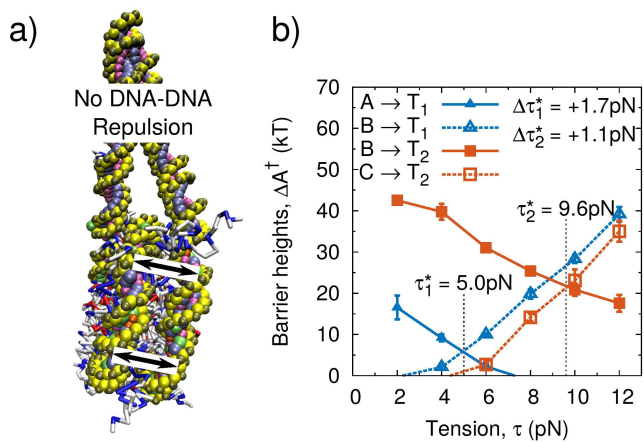}
    \caption{%
    a) Schematic of proposed model with DNA-DNA repulsion removed.
    b) Resulting tension-dependent free energy barriers for 601 positioning sequence.
    $\Delta \tau_1^*$ and $\Delta \tau_2^*$ represent change relative to complete model.
    Error bars represent standard deviation across three independent simulations.
    }
    \label{fig:noelec}
\end{figure}

We first investigate the origins of the tension-dependent mechanical response by exploring the role of DNA-DNA electrostatic repulsion on the stability of the nucleosome structure. 
Past theoretical work \cite{Kulic2004,Mollazadeh-Beidokhti2012} has suggested that DNA-DNA repulsion within the nucleosome is central to its tension-dependent response. 
Other studies, however, have observed that DNA-DNA electrostatic repulsion is unimportant and that the correct response can be achieved by accounting for the tension-dependent orientation of the free DNA ends\cite{Sudhanshu2011}.
Since our proposed model explicitly includes both contributions, we can directly evaluate the importance of DNA-DNA repulsion on nucleosome unwrapping.
To examine this effect, we disable DNA-DNA electrostatic repulsion in our model between base-pairs separated by more than 20 base pairs.
Only disabling electrostatics between distant regions of DNA was necessary to avoid implicitly lowering the persistence length of DNA by neglecting Coulombic interactions between neighboring base-pairs.
All electrostatics responsible for DNA-histone affinity however, remain intact.

Our results are summarized in Figure~\ref{fig:noelec}a,b.
As anticipated\cite{Kulic2004}, removal of DNA-DNA repulsive interactions has a greater impact on the outer DNA loop ($\Delta \tau_1^* = + 1.7$pN) than on the inner DNA loop ($\Delta \tau_2^* = + 1.1$pN).
However in the absence of DNA-DNA repulsions, the qualitative features of the tension-dependent response remain unchanged.
These results indicates that while DNA-DNA repulsions play a role in nucleosome disassembly, they are not primarily responsible for the two unwrapping steps observed in experiments.

\begin{figure}

    \begin{overpic}[width=3.33in]
    {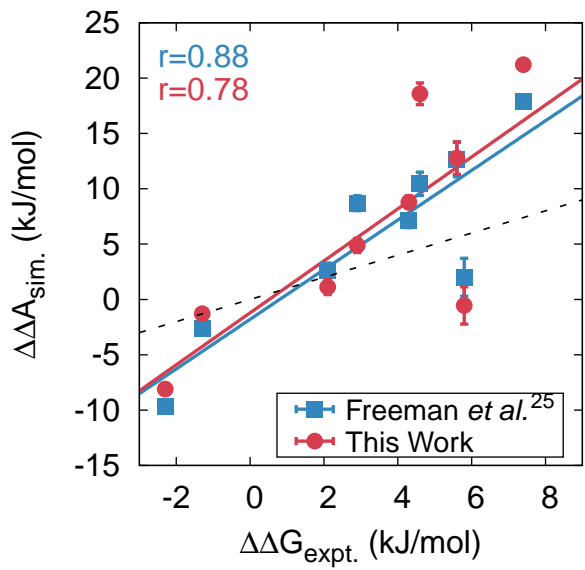}
    \put (54,30){\sffamily Freeman \textit{et al.}~\cite{Freeman2014a}}
    \put (54,25){\sffamily This Work}
    \end{overpic}
    \caption{%
    Sequence-dependent binding free energies.
    Squares denote model proposed by Freeman \textit{et al.}~\cite{Freeman2014a} (obtained at 300K and 150mM ionic strength).
    Circles denote model proposed in this work, obtained at 277K and vanishing ionic strength (for consistency with Ref.~\cite{Thastrom2004a}).
    Despite differing solution conditions and DNA-Protein interactions, both models reproduce the \textit{relative} binding free energies of nucleosome formation.
    The DNA sequences used here are given in Ref.~\cite{Freeman2014a}.
    }
    \label{fig:ti}
\end{figure}

We next examine the impact of DNA sequence on the relative binding free energies of nucleosome formation.
Optical trapping experiments could in principle be used to probe the sequence-dependent energies within the nucleosome, but recent literature studies have been limited to the 601 positioning sequence \cite{Mihardja2006,Kruithof2009,Sheinin2013} and slight variations \cite{Ngo2015}.
Instead, competitive reconstitution assays are the dominant experimental technique for characterization of sequence-dependent relative binding free energies\cite{Thastrom2004a,segal2006genomic}.
To compare model predictions to these experiments, we use the technique employed by Freeman \textit{et al.}~\cite{Freeman2014a}, where the relative binding free energies of different DNA sequences are assessed computationally using alchemical transformations and thermodynamic integration (see Materials and Methods).
A comparison of predicted and experimental free energies, shown in Figure~\ref{fig:ti}, indicates that,
as with previous work\cite{Freeman2014a}, the model adopted here accurately reproduces the binding free energies of many different sequences.
In general, the key predictor of binding free energy is the sequence-dependent shape of the DNA molecule (i.e. minor groove widths and intrinsic curvature).
Sequences that bind strongly (low $\Delta\Delta A$) posses periodic sequence motifs (e.g. TA base steps) that impart a shape that favorably ``fits'' underlying histone structure\cite{segal2006genomic}.
In contrast, weakly binding sequences (large $\Delta\Delta A$) do not posses these periodic motifs.

\begin{figure}
    \includegraphics[width=3.33in]{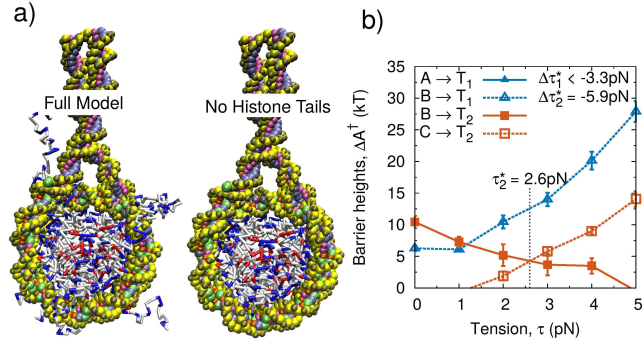}
    \caption{%
    a) Schematic of proposed model with histone tails removed.
    b) Resulting tension-dependent free energy barriers for 601 positioning sequence.
    $\Delta \tau_1^*$ and $\Delta \tau_2^*$ represent change relative to model with intact histone tails.
    Error bars represent standard deviation across three independent simulations.
    }
    \label{fig:notail}
\end{figure}

In addition to DNA sequence, the modification of histone tails is widely considered to be the single most important determinant of chromatin structure \cite{Jenuwein2001}.
Methylated and acetylated histones are enriched at promoters of highly expressed genes and are thought to play a role in the strong positioning of certain nucleosomes \citep{kurdistani2004mapping,barski2007high}.
Histone tails are central to nucleosome-nucleosome interactions and their modification has important implications on chromatin's three-dimensional structure\citep{zhou2011charting}.
Experiments\cite{Brower-Toland2005} have also established that removal of histone tails has a significant impact on the stability of the nucleosome.

To examine the role of histone tails on nucleosome stability at a molecular level, we return to our earlier analysis and calculate the tension-dependent free energy of nucleosome unwrapping. Our results can be compared to the optical trapping experiments of Brower-Toland \textit{et al.}~\cite{Brower-Toland2005}, where arrays of 17 nucleosomes were disassembled for different histone tail modifications, including complete removal via trypsin digest or post-translationally via acetylation.
In the model, we perform this trypsin digest \textit{in silico} (see Figure~\ref{fig:notail}a), and calculate the resulting tension-dependent response (Figure~\ref{fig:notail}b). Our results indicate that removal of histone tails significantly destabilizes the outer loop, causing the ``Partially Wrapped'' (c.f. Figure~\ref{fig:pull-snapshots}b) state to be energetically favored at zero tension.
Additionally, histone tail removal shifts the inner loop unwrapping force, $\tau_2^*$, to $2.6$pN, $5.9$ pN lower than when histone tails were intact (c.f. Figure~\ref{fig:barrier}).
Our results are consistent with measurements that reported a 3 pN reduction in ``peak force'' upon tail removal and a 60\% reduction in the outer turn length\cite{Brower-Toland2005}.
In this case, quantitative agreement is not expected since these experiments lacked sufficient spatial resolution to resolve the individual release of the outer DNA turn.

\section*{Conclusion}

In this work we have demonstrated that a molecular-model of the nucleosome, composed of two coarse-grained models of DNA and proteins\cite{Freeman2014a,freeman2014coarse, Li2011}, can be combined parameter-free to accurately reproduce the tension-dependent response of nucleosome unwrapping.
This model quantitatively reproduces the unwrapping forces observed in experiments\cite{Mihardja2006} and the barrier heights predicted by prior theoretical studies \cite{Sudhanshu2011}.
We then demonstrated that this model can be used to examine, without adjustment, the role of subtle phenomena in nucleosome formation such as DNA-DNA Coulombic repulsion, DNA-sequence, and histone tail modifications.

Our proposed approach opens up a new avenue for theoretical examinations of nucleosome stability.
As a first step, this model can aid the interpretation of recent optical pulling experiments where nucleosome are subjected to torque \cite{Sheinin2013}, and is suitable to examine subtle features within the nucleosome such as asymmetric unwrapping \cite{Ngo2015}.
Yet the potential of our approach extends beyond nucleosome pulling experiments, and can begin to elucidate many unsolved questions within chromatin biophysics.
How does the methylation of specific histone tails (and not others) enhance the positioning of certain nucleosomes?
What are the free energies of different folded chromatin structures, and how do histone modifications effect this energy landscape?
What is the role of DNA sequence on these processes, and do certain DNA sequences dispose chromatin to different ``folds''?
The approach presented in this work represents an important step towards answering these questions.

\section*{Methods} \label{sec:methods}
The model adopted in this work relies on a coarse-grained model of DNA \cite{freeman2014coarse,Freeman2014a} and proteins \cite{Li2011}, which are combined to represent the nucleosome. Both models were developed independently, but they are implemented at the same level of description, thereby facilitating their concerted use. Specifically, for DNA we use the 3SPN coarse-grained representation, where each nucleotide is described by three force sites located at the phosphate, sugar and base\cite{Knotts2007,Sambriski2009,Hinckley2013,freeman2014coarse}.
For the histone proteins, we use the ``Atomistic-Interaction based Coarse-Grained model'' (AICG), where the protein is represented by one site per amino acid located at the center of mass of the sidechain\cite{Li2011}.

Interactions between the 3SPN2.C and AICG models included electrostatic and excluded volume effects.
Phosphate sites with 3SPN were assigned a charge of $-0.6$ as described previously\cite{Hinckley2013}.
Each protein site was given the net charge of that residue at physiological pH (i.e. +1 for Arg, Hys and Lys; -1 for Asp and Glu, 0 for others).
As with prior work\cite{Freeman2014a}, the effective charge of interactions between DNA and protein sites was scaled by a factor of 1.67 to bring the local charge of the phosphates back to -1.
We note that DNA-Protein interactions in this work differ slightly from those employed by Freeman \textit{et al.}~\cite{Freeman2014a} where, in addition to electrostatics, a small Lennard-Jones attraction was added between all DNA and protein sites.
The strength of this attraction was very weak ($\epsilon_{Pro-DNA} = 0.25 kJ/mol$) and was originally included to reduce fluctuations within the nucleosome structure. Here we demonstrate that this weak interaction is unnecessary; by omitting it, both the relative and absolute formation free energies of the nucleosome can be reproduced. The combined model is effectively parameter-free: both the model of DNA and Protein are included as originally proposed without any additional terms.
Electrostatic forces are introduced at the level of Debye-H\"{u}ckel theory.
All simulations were performed in the canonical ensemble using a Langevin thermostat and 150mM ionic strength.

As an initial condition, we combine the 1KX5 crystal structure \cite{Davey2002} of the nucleosome core particle with a proposed configuration of exiting DNA \cite{Hussain2010,Meyer2011} to form a 223 base-pair structure, with 147 base-pairs bound to the histone proteins and 38 flanking bases on each side.
When using the 601 positioning sequence\cite{Lowary1998}, the flanking bases were chosen as polyA.
This configuration was only used as the initial configuration, and no information from either structure was directly encoded into the nucleosome model.

To extract the tension-dependent free energy surface, two constraints were applied to the nucleosomal model.
First, a constant force (i.e. tension) was applied to each end of DNA in order to mimic the experimental setup of optical-trapping experiments.
Then, harmonic constrains were applied to the end-to-end extension of the DNA molecule, and umbrella sampling was performed to determine the free-energy as a function of DNA extension\cite{Kumar1995, Kastner2011}.
This simulation framework results in free energy ``surfaces'' that are not truly continuous for different values of tension.
They are instead a compilation of two-dimensional ``curves'' that are plotted co-currently to construct the ``surface'' presented in Figure~\ref{fig:tension-sep}.

The relative free energy of binding for different DNA sequences ($\Delta\Delta A$) was calculated as described in detail previously \cite{Freeman2014a}.
Briefly, a thermodynamic cycle was defined that represents the relative sequence-dependent free energy of nucleosome formation, $\Delta \Delta A$, as the difference between the free energy difference of two DNA sequences in the bulk, $\Delta A_{bulk}$, and bound to the histone proteins, $\Delta A_{bound}$ (i.e. $\Delta\Delta A = \Delta A_{bulk} - \Delta A_{bound}$);
$\Delta A_{bulk}$ and $\Delta A_{bound}$ are determined by thermodynamic integration.
The DNA sequences analyzed are given explicitly in the original paper.

\begin{acknowledgement}
This work was supported by NIST through the Center for Hierarchical Materials Assembly (CHiMaD). 
Computational resources were provided by the Midway computing cluster at the University of Chicago and the University of Wisconsin--Madison Center for High Throughput Computing.
The authors gratefully acknowledge Professor Andrew Spakowitz for helpful discussions and Ralf Everaers and Sam Meyer for providing the configuration of the 223 base-pair nucleosome.
\end{acknowledgement}

\begin{suppinfo}

\end{suppinfo}
\clearpage
\providecommand{\latin}[1]{#1}
\providecommand*\mcitethebibliography{\thebibliography}
\csname @ifundefined\endcsname{endmcitethebibliography}
  {\let\endmcitethebibliography\endthebibliography}{}

\end{document}